\documentclass[preprint,notoc]{JHEP3}
\usepackage{epsfig}

\def\to{\rightarrow}

\def\te{\tilde e}

\def\tu{\tilde u}

\def\tb{\tilde b}

\def\td{\tilde d}

\def\tst{\tilde t}
\def\ttau{\tilde \tau}
\def\tmu{\tilde \mu}
\def\tg{\tilde g}
\def\tnu{\tilde\nu}
\def\tell{\tilde\ell}

\def\tw{\widetilde W}
\def\tz{\widetilde Z}

\title{Viable models with non-universal gaugino mediated supersymmetry 
breaking}

\author{Howard Baer, Csaba Bal\'azs, Alexander Belyaev\footnote
{On leave of absence from Nuclear Physics Institute, Moscow State University.} 
\\ Department of Physics, Florida State University\\ 
Tallahassee, FL, USA 32306\\
E-mail: \email{baer@hep.fsu.edu},\email{balazs@hep.fsu.edu},
        \email{belyaev@hep.fsu.edu}}
\author{Radovan Derm\' \i \v sek \\
Department of Physics, The Ohio State University, \\
Columbus, OH 43210, USA    \\
E-mail: \email{dermisek@pacific.mps.ohio-state.edu}}
\author{Arash Mafi\\
Department of Physics, University of Arizona\\
Tucson, AZ 85712\\
E-mail: \email{mafi@physics.arizona.edu}}
\author{Azar Mustafayev
\\ Department of Physics, Florida State University\\ 
Tallahassee, FL, USA 32306\\
E-mail: \email{mazar@hep.fsu.edu}}

\preprint{FSU-HEP-020408\\ OHSTPY-HEP-T-02-004}

\abstract{Recently, extra dimensional SUSY GUT models have been
proposed in which compactification of the extra dimension(s) 
leads to a breakdown of the gauge symmetry and/or supersymmetry.
We examine a particular class of higher-dimensional models exhibiting
supersymmetry and $SU(5)$ or $SO(10)$ GUT symmetry. SUSY breaking occurs 
on a hidden brane, and is communicated to the visible brane via gaugino 
mediation. Non-universal gaugino masses are developed at the 
compactification scale
as a consequence of a restricted gauge symmetry on the hidden 
brane. In this case, the compactification scale is at or slightly below
the GUT scale. We examine the parameter space of such models 
where gaugino masses are related due to a Pati-Salam symmetry on the hidden 
brane. We find limited but significant regions of model parameter space where
a viable spectra of SUSY matter is generated. 
Our results are extended to the more general case of three independent
gaugino masses; here we find that large parameter space regions open
up for large values of the $U(1)$ gaugino mass $M_1$.
We also find the relic density
of neutralinos for these models to be generally below
expectations from cosmological observations, thus leaving room
for hidden sector states to make up the bulk of cold dark matter.
Finally, we evaluate the branching fraction $BF(b\to s\gamma )$ and
muon anomalous magnetic moment $a_\mu$.}


\begin{document}

\section{Introduction}
\label{sec:introduction}

Introducing supersymmetry (SUSY) to solve the hierarchy problem 
gives rise to a puzzle of identifying the way it 
is broken. 
Several mechanisms for SUSY breaking have been invented.
Although at present there is no compelling reason to prefer one 
to the other, we know that the soft SUSY breaking terms 
in the minimal supersymmetric standard model (MSSM) must have a 
very special form. 
In order to avoid unacceptably large flavor violation, 
the masses of squarks and sleptons must be degenerate to a
high precision\footnote{Other solutions to the SUSY flavor problem
such as decoupling or alignment are generally more awkward to implement
in models.}. 
Similarly the trilinear couplings must be proportional to the Yukawa matrices. 
Therefore it is intriguing to explore scenarios which naturally 
produce soft SUSY breaking parameters with these desired properties. 
Examples of such scenarios are gauge mediation~\cite{gMSB}, 
anomaly mediation~\cite{anomalyMSB1, anomalyMSB2}, and gaugino 
mediation~\cite{ginoMSB1, ginoMSB2}.

In the gaugino mediation scenario, the MSSM matter fields are 
localized on a 3-brane --- the ``matter" brane 
embedded in higher dimensional spacetime --- 
while gauge fields can propagate in the bulk of the extra dimensions. 
Supersymmetry is broken on a separate ``hidden sector'' brane which 
is separated from the matter brane in extra dimensions. 
Gauginos can 
directly couple to the source of SUSY breaking to obtain nonzero masses, 
while masses of squarks and sleptons 
and trilinear couplings are suppressed due to the 
spatial separation of the branes.
Thus, just below the compactification scale 
$M_c$\footnote{$M_c \sim 1/R$, R represents the size of extra 
dimension(s).}, the effective 4-dimensional theory 
is the MSSM and in the minimal version  gaugino masses
are the only non-negligible soft SUSY breaking 
parameters.\footnote{Soft SUSY breaking Higgs masses can also
be generated if Higgs fields propagate in the bulk.} 
Scalar masses and trilinear couplings receive large 
contributions from gaugino masses through the renormalization group (RG) 
running between $M_c$ and the electroweak (EW) scale.
These contributions are flavor blind and therefore the 
resulting soft SUSY breaking terms at the EW scale 
cause only a modest flavor violation coming from the Yukawa couplings.

In earlier works on gaugino mediation~\cite{ginoMSB1, ginoMSB2},
the compactification scale was assumed to be at or above
the GUT scale, 
in order to preserve the success of gauge coupling unification. 
Therefore all gaugino masses are equal at $M_c$.
For $M_c = M_{GUT}$, however, this scenario predicts 
the lightest stau $\ttau_1$ to be the lightest supersymmetric 
particle (LSP), which violates cosmological bounds on the existence
of stable charged or colored relics from the Big Bang in models
which conserve $R$-parity.
The situation is different for $M_c > M_{GUT}$.
In this case the additional RG running between $M_c$ and $M_{GUT}$
has to be considered~\cite{ginoMSB2}. This 
generates non-zero scalar masses and trilinear couplings at the GUT scale;
most importantly the stau mass gets a positive
contribution which eventually can make the $\ttau_1$ heavier than the
lightest neutralino $\tz_1$. This removes the unpleasant
charged LSP feature of the scenario with
$M_c = M_{GUT}$\cite{ginoMSB2,gaugino_Baer}.
In models which assume a simple GUT beyond $M_{GUT}$,
the values of $\tan\beta$ ($\sim 35-50$) consistent with
Yukawa coupling unification do not
match the $\tan\beta$ values ($\sim 10-25$) predicted by the presence
of a small bilinear soft SUSY breaking
parameter $B_0$\cite{gaugino_Baer}.

There has been recent interest in constructing GUT models 
in higher dimensional spacetimes \cite{kawamura, 
hall_nomura, other_models, so10in5d, so10in6d}.
Orbifold compactification of the extra dimensions is 
an elegant way of achieving GUT symmetry breaking and 
the doublet triplet splitting of Higgs fields. 
Proton decay due to dimension 5 operators that is a 
serious problem in 4-dimensional GUT models \cite{proton},
can be naturally suppressed in these models.
The common feature of these models is the existence of a 
brane or several branes at orbifold fixed points on which the 
gauge symmetry is restricted to be a subgroup of the GUT symmetry. 
This results in an effective 4-dimensional theory with 
gauge symmetry given by an intersection of gauge 
symmetries on orbifold fixed points. 

Positive features of GUTs, like charge quantization or gauge 
coupling unification, can be preserved. 
If matter multiplets live in the bulk or are localized on a brane 
which respects the GUT symmetry, then they come in complete GUT 
multiplets. Furthermore, if the compactification scale is not 
far from $M_{GUT}$, the gauge couplings still unify. 
The gauge couplings run from the
EW scale to the compactification scale with the ordinary MSSM
$\beta$ functions. They almost unify at $M_c$ which is below but   
very close to the conventional GUT scale. Beyond $M_c$, the gauge
couplings run according to the power-law rule~\cite{DDG} 
due to the heavy KK modes that do not fill degenerate GUT 
multiplets; they ultimately unify at a scale $M_*$, the cutoff 
scale of the model. 

The existence of a brane with restricted gauge symmetry can 
play an important role for gaugino mediation. 
If SUSY is broken on a brane with restricted gauge symmetry, 
non-universal gaugino masses are generated. For example,
if using proper boundary conditions
$SU(5)$ is broken on a brane down to the standard model (SM), 
non-universal gaugino masses $M_1, M_2, M_3$ can be 
generated on this brane~\cite{hall_nomura}. 

Even more interesting is the situation for $SO(10)$ models in 
higher dimensions~\cite{so10in5d, so10in6d} which can 
contain branes with gauge symmetries being different subgroups of $SO(10)$. 
Using proper boundary conditions, fixed 
branes with Pati-Salam $SU(4) \times SU(2)_L \times SU(2)_R$, 
Georgi-Glashow $SU(5) \times U(1)$, or 
flipped $SU(5)^\prime \times U(1)^\prime$ gauge symmetries can be obtained. 
If gauginos get masses on these branes, the 
gauge symmetry relates gaugino masses of the MSSM at the 
compactification scale. In the case of Pati-Salam gauge symmetry, 
gaugino masses $M_2$ and $M_3$ are free parameters while the 
$M_1$ is given by a linear combination 
of $M_2$ and $M_3$\cite{so10in5d}.
 
The compactification scale in these models is below the GUT scale 
and  the boundary conditions with negligible 
squark and slepton masses and trilinear couplings are realized at this scale. 
Therefore the cure of Ref. \cite{ginoMSB2} to the 
stau LSP problem doesn't apply here. However, in this case the
non-universal gaugino masses may help to obtain viable SUSY 
spectra with a neutralino LSP, at least in some restricted 
regions of model parameter space. 
Non-universal gaugino masses can also be very helpful in 
order to obtain Yukawa coupling unification for models with
a positive $\mu$ term\cite{Yukawa_unif}.  

We emphasize that extra dimensional gauge theories are not the only way 
to get the MSSM at the GUT scale with the non-universal gaugino masses 
as the only sources of the soft SUSY breaking.  
For example, 
one may imagine deconstructing the 5-dimensional models with non-universal 
gaugino masses (for discussion of deconstructed universal gaugino mediation 
see~\cite{Csaki:2001em}).
In addition, models of no-scale supergravity can also provide 
us with the desired boundary conditions i.e. 
vanishing scalar masses and trilinear
couplings and non-zero gaugino masses\cite{dimitri}. 
In a no-scale locally supersymmetric
GUT, a non-trivial choice of the couplings between the vector supermultiplets
and the chiral supermultiplet can easily result in non-universal gaugino masses
at the GUT scale~\cite{Ellis:1985jn}.  
Similar results can also be obtained in type I string constructions, see
for example~\cite{king:2001}.

In this paper, we present the allowed region of SUSY parameter space 
and study the supersymmetric particle spectrum for 
the case where the bino mass $M_1$ is 
related to $M_2$ and $M_3$ by the Pati-Salam symmetry. 
In this case we {\it do} find regions of
parameter space leading to a viable SUSY particle spectrum although the
allowed regions turn 
out to be rather tightly restricted. 
We also evaluate the relic density of
neutralinos for these models; we find it in general to be somewhat
below expectations from cosmological measurements due to the presence of a 
non-negligible higgsino component to the LSP.
This is a positive feature of such models, since one may expect 
additional cold dark matter (CDM) states associated for 
instance with the hidden brane.  
We evaluate as well the branching fraction $BF(b\to s\gamma )$ and
muon anomalous magnetic moment $a_\mu =\frac{(g-2)_\mu}{2}$.
We also study the effect of adding non-zero soft SUSY breaking Higgs 
masses on the allowed regions of parameter space; typically, 
these Higgs masses do not ameliorate the situation. 
Finally, we examine the case of
completely independent non-universal gaugino masses 
at the compactification scale. In this case, we find that regions of
parameter space leading to viable spectra open up
for large values of the bino mass $M_1$. 
We show sample spectra for these models; they typically lead to
spectra with nearly degenerate $\tw_1$ and $\tz_1$, since the LSP 
turns out to be
either wino-like (as in AMSB models) or higgsino-like.

This paper is organized as follow. In Sec. \ref{sec:model}, we 
review some details of  extra dimensional SUSY GUT models with
gaugino mediated SUSY breaking, and their associated parameter space.
In Sec. \ref{sec:PS}, we examine models where $M_1$ is determined
in terms of $M_2$ and $M_3$ by the Pati-Salam gauge symmetry, and
show that there exists parameter space leading to viable models, although
its extent is somewhat limited.
In Sec. \ref{sec:general}, we present our
results for completely independent gaugino masses. These models
emphasize regions of parameter space with large $U(1)_Y$ gaugino mass $M_1$,
and can frequently lead to spectra with a wino-like LSP.
Our conclusions are presented in Sec. \ref{sec:concl}. 

\section{Non-universal Gaugino Mediation}
\label{sec:model}

In this section, we will review the basic features of GUT models in 
higher dimensional spacetimes. We start with the simple example of an 
$SU(5)$ 
GUT in five dimensions (for more detailed discussion, see \cite{hall_nomura}).
The fifth dimension is compactified 
on an $S^1/(Z_2\times Z^\prime_2)$ orbifold 
where $S^1$ is a circle of radius R defined with a 
periodic coordinate $0\leq y < 2\pi R$.
$Z_2$ identifies opposite points on the circle $y \to - y$ and  
$Z^\prime_2$ identifies opposite points 
with respect to $y^\prime = y - \pi R/2$, i.e. $y^\prime \to - y^\prime$. 
The resulting orbifold has two 
inequivalent fixed points $O$ at $y = 0$ and $O^\prime$ at 
$y = \pi R/2$ ($y^\prime = 0$).
There are fixed branes (4-dimensional Minkowski spaces) at these points. 
Let's assume that the gauge supermultiplet  
and two Higgs hypermultiplets in {$\bf 5$} and {$\bf \bar 5$} 
representations propagate in the bulk. In 
five dimensions, $\cal{N}$~=~1 
supersymmetry
is generated by 8 supercharges and is equivalent to $\cal{N}$~=~2
supersymmetry in four dimensions. 
Using 4 dimensional superfield notation, the 5-dimensional gauge 
supermultiplet can be written as $(V, \Sigma)$ where $V$ is 
a 4-dimensional gauge supermultiplet and 
$\Sigma $
is a chiral supermultiplet, both in the adjoint 
(24-dimensional) representations of $SU(5)$.
Similarly Higgs hypermultiplets that transform as 
{$\bf 5$} and {$\bf \bar 5$} can be written as $(H_5, 
H^c_5)$ and  $(H_{\bar 5}, H^c_{\bar 5})$ where $H_5$ 
and $H^c_5$ ($H_{\bar 5}$ and $H^c_{\bar 5}$) are 
4-dimensional chiral supermultiplets in the $\bf 5$ and 
$\bf \bar 5$ ($\bf \bar 5$ and $\bf 5$) representations of $SU(5)$.

A generic field
$\phi(x_\mu,y)$ in the 5-dimensional bulk is identified by its
transformations under the $Z_2$ and $Z^\prime_2$ parities $P=\pm$
and $P^\prime=\pm$, respectively; 
$\phi (-y) = P \phi (y)$ and $\phi (-y^\prime) = P^\prime \phi (y^\prime )$. 
Only those fields with parity $(+ , + )$ under $Z_2\times Z^\prime_2$ 
have massless 
Kaluza-Klein modes in the resulting 4-dimensional effective theory. 
Invariance of the 5-dimensional gauge interactions under 
$Z_2$ requires a relative minus 
sign between the parity transformations of $V$ and $\Sigma $ and 
similarly between $H$ and $H^c$ (in both $\bf 
5$ and $\bf \bar 5$). 
$Z_2$ parities of $V$, $H_5$ and $H_{\bar 5}$ are taken to be 
$+ $ and those of $\Sigma $, $H^c_5$ and 
$H^c_{\bar 5}$ are taken to be $-$. Therefore the 
orbifold compactification breaks $\cal{N}$~=~2 supersymmetry 
in four dimensions down to  $\cal{N}$~=~1.
  
Using a non-trivial assignment of $Z_2^\prime$ parities 
$P^\prime= \, \rm diag \, (-1,-1,-1,+1,+1)$ 
acting on $\bf 5$, the SU(5) gauge symmetry is restricted to the 
standard model gauge symmetry 
$SU(3)_C \times SU(2)_L \times U(1)_Y \equiv G_{SM}$ on the $O^\prime$ brane. 
At the same time $P^\prime$ 
splits Higgs doublets from their color triplet partners. As a result 
only gauge fields of the minimal supersymmetric standard model 
and two Higgs doublets $H_u$ and $H_d$ (those contained in 
$H_5$ and $H_{\bar 5}$) are (+,+) 
fields and therefore will be 
present at the massless level in the resulting 4-dimensional effective theory.

The third generation matter fields can be localized on the $O$ brane. 
Since the gauge symmetry on the 
$O$ brane is SU(5), matter fields must come in complete SU(5) multiplets. 
Therefore we can have usual 
SU(5) $b - \tau$ Yukawa unification although the 4-dimensional 
theory is just the MSSM.

This setup with two separated fixed branes is very 
suitable for gaugino mediation. 
Let's assume that SUSY is
broken by the vacuum expectation value of a gauge
singlet chiral superfield $X$ that is localized on the $O^\prime$ brane,
\begin{equation}
\label{eq:X_vev}
\langle X \rangle=\theta^2 F_X.
\end{equation}
$X$ can couple directly to the gauge fields on the $O^\prime$ brane through
the ultraviolet scale suppressed terms
\begin{equation}
{\cal L}_{O^\prime} \supset
\int{d^2\theta}\Big(\lambda_3 \frac{X}{M^2_*}W^{i\alpha}  
W^i_\alpha
+ \lambda_2 \frac{X}{M^2_*}W^{j\alpha}W^j_\alpha 
+ \lambda_1 \frac{X}{M^2_*}W_Y^{\alpha}W_{Y \alpha} 
+ H.c.\Big),
\end{equation}
where index $i(j)$ runs over the gauge fields of the
$SU(3) (SU(2))$ gauge group.
This will give masses to gauginos of $SU(3)$, $SU(2)_L$ and $U(1)_Y$
separately,
\begin{equation}
\label{eq:gauginos_su5}
M_i=\frac{\lambda_i F_X M_c}{M^2_*}, \qquad i = 1,2,3
\label{eq:genM}
\end{equation}
which corresponds to the fact that the gauge symmetry 
on the $O^\prime$ brane is restricted to $G_{SM}$.
The factor $M_c$ (the compactification scale, $M_c \sim 1/R$) 
comes from the wave function normalization
of the 4-dimensional gaugino fields.
Similarly the $\mu$ and $B_\mu$ terms ($B_\mu =B\mu$) can be generated:
\begin{equation}
{\cal L}_{O^\prime} \supset
\int{d^4\theta} \Big( \frac{X^\dagger}{M^2_*}
H_u H_d + \frac{X X^\dagger}{M^3_*} H_u H_d + H.c. \Big),
\end{equation}
leading to
\begin{equation}
\mu\sim \frac{F_X M_c}{M^2_*} , \qquad B_\mu\sim \frac{F_X^2  M_c}{M^3_*}.
\end{equation}
Finally soft SUSY breaking masses for Higgs scalars can be generated:
\begin{equation}
{\cal L}_{O^\prime} \supset
\int{d^4\theta} \Big( \frac{X X^\dagger}{M^3_*}
H_u H_u^\dagger + \frac{X X^\dagger}{M^3_*} H_d H_d^\dagger  \Big),
\end{equation}
giving
\begin{equation}
m^2_{H_u} \sim m^2_{H_d} \sim \frac{F_X^2  M_c}{M^3_*}.
\end{equation}

To summarize the results of this model, 
in the effective 4-dimensional theory at the massless level we 
have MSSM gauge fields and two Higgs doublets.
At the compactification scale non-universal gaugino masses 
and the $\mu$ term are generated 
together with $B_\mu$ term and soft SUSY breaking Higgs masses. 
Soft SUSY breaking masses of squarks 
and sleptons and trilinear couplings are negligible at $M_c$. 
Furthermore, Yukawa couplings of $b$ and $\tau$ unify at $M_c$.

The fact that in some sectors of the theory
there are GUT relations between parameters and in others there are no
relations is clearly a consequence of having branes with
grand unified gauge symmetry and with restricted gauge symmetry.
The results of the above model can also be easily modified by 
different arrangements of fields living in the bulk and 
those localized on branes. 
For example if the third generation matter fields are 
not localized on the brane with GUT 
symmetry (they can be localized on the brane with restricted gauge
symmetry or can come from different multiplets living 
in the bulk), the Yukawa unification is not 
predicted.\footnote{This can be actually very useful for treating 
first two 
generation, since the unification of Yukawa couplings of 
first two generation fermions is excluded by experimental 
observations.} Similarly if the Higgs fields do not live 
in the bulk and are not localized on the brane where SUSY is 
broken, the soft SUSY breaking Higgs masses are not generated. 
In this case the $\mu$ term is not generated either;
however, it can be generated within the effective 
4-dimensional theory~\cite{4d_mu}. Therefore, generating gaugino masses 
is not necessarily connected with either generating 
Higgs masses or Yukawa unification.   

A very interesting situation can happen in models with 
$SO(10)$ GUT symmetry in higher dimensional 
spacetimes~\cite{so10in5d, so10in6d}. 
There are more ways to break $SO(10)$ symmetry down to $G_{SM}$ and therefore 
branes on which the gauge symmetry is restricted to 
different subgroups of $SO(10)$ can appear in a model.
Splitting of Higgs doublets and triplets can be achieved 
if the gauge symmetry on a brane is 
restricted to $SO(6) \times SO(4)$ 
which is isomorphic to the Pati-Salam gauge symmetry 
$SU(4) \times SU(2)_L \times SU(2)_R$. 
In a similar way as in Eqs.~(\ref{eq:X_vev})~-~(\ref{eq:gauginos_su5}), 
masses of gauginos of $SO(6)$ and $SO(4)$
gauge groups can be generated on this brane separately~\cite{so10in5d},
\begin{equation}
\label{eq:gauginos}
M_6=\frac{\lambda_6^\prime F_X M_c}{M^2_*}, \qquad
M_4=\frac{\lambda_4^\prime F_X M_c}{M^2_*}.
\end{equation}
The masses of gauginos
of the MSSM $(M_1, M_2, M_3)$ are then given as
\begin{equation}
M_1=\frac{2}{5}M_6+\frac{3}{5}M_4,\qquad M_2=M_4,\qquad M_3=M_6.
\label{Eq:PatiSalamMi}
\end{equation}
The special form of $M_1$ is related to the way in which the 
generator of hypercharge
is expressed as a linear combination of the generator of 
$B-L$ from $SO(6)$ and the generator of $t_{3R}$ from 
$SO(4)$, 
namely
$Y=\sqrt{\frac{2}{5}}(B-L)-\sqrt{\frac{3}{5}}t_{3R}$ 
(for detailed discussion see~\cite{so10in5d})~\footnote{Similarly, if 
gaugino masses are generated on a brane with flipped $SU(5)^\prime \times 
U(1)^\prime$ symmetry, we have relation $M_2 = M_3 = M_{5^\prime}$ and 
$M_1$ is a linear 
combination of $M_{5^\prime}$ and $M_{1^\prime}$. Therefore gaugino masses 
are given by two independent parameters $M_1$ and $M_2 = M_3$.}.

As in the case of $SU(5)$, Yukawa unification may or may not be present, 
depending on the way matter 
fields are introduced in the model. If the third generation matter 
fields are localized on a brane with $SO(10)$ 
symmetry, $t-b-\tau$ Yukawa unification is predicted. 
However, matter fields can be localized on a brane with 
$SU(5)\times U(1)$ gauge symmetry in which case only 
$b-\tau$ Yukawa unification should be expected at $M_c$. And of 
course models can be built without having any kind of Yukawa unification. 
Similarly soft SUSY breaking Higgs masses 
can be generated. If they are generated on the same brane as gaugino masses, 
$SO(6)\times SO(4)$ symmetry guarantees 
that $m_{H_u}^2 = m_{H_d}^2$ at the compactification scale.

\section{Pati-Salam scenario}
\label{sec:PS}

In this section, we investigate the parameter space of gaugino
mediated SUSY breaking models where gaugino masses are constrained
by the Pati-Salam (PS) symmetry to obey
\begin{equation}
M_1={3\over 5} M_2+ {2\over 5}M_3
\end{equation}
at the compactification scale. 
Scalar masses and trilinear soft SUSY breaking terms
are restricted to be $m_0\simeq A_0\simeq 0$. In minimal 
gaugino mediation the bilinear $B_0$ parameter is also $\simeq 0$,
which leads to a predicted value of $\tan\beta$. Here, we will work with
a somewhat more general format, taking $\tan\beta$ as an independent
parameter and determining the value of $B$ at the weak scale by imposing 
minimization conditions on the scalar potential, so that
radiative EW symmetry breaking occurs (REWSB). Note that this procedure
also determines the magnitude, but not the sign, of the superpotential
Higgs mass term $\mu$.
 
Our goal is to delineate the allowed regions of the parameter 
space in this scenario, and to find the regions that can be explored by 
near future experiments.
The parameter space of the model is then given by
\begin{equation}
M_2(M_c), ~ M_3(M_c), ~ \tan\beta, ~ sign(\mu), ~ M_c
\end{equation}
as the independent parameters. The sparticle spectrum will depend
only logarithmically on the value of $M_c$, which nonetheless must be near 
the GUT scale. Hence, we adopt the choice
$M_c = 1 \times 10^{16}$ GeV, {\it i.e.} just below the 
usual unification scale of $M_{GUT}\simeq 2\times 10^{16}$ GeV.

We employ an updated version of ISAJET version 7.58\cite{isajet} which allows
for sparticle spectra predictions for negative values of the gaugino mass
$M_3$ (as well as $M_2$ and $M_1$ as in earlier versions). 
Briefly, the subprogram 
ISASUGRA (which is a part of ISAJET) begins with weak scale values of
gauge and Yukawa couplings, and evolves up in energy until the
unification scale where $g_1=g_2$ is identified. 
At $M_{GUT}$ (or other intermediate scales), 
the soft SUSY breaking boundary conditions can be imposed, and evolution of
the set of 26 coupled RGEs of the MSSM is calculated down to the weak scale,
where the 1-loop effective potential is minimized at an optimized scale
choice which reproduces to within $1-2$~GeV the light scalar Higgs mass
$m_h$ as predicted by the FeynHiggsFast program\cite{FHF}.
Yukawa couplings are updated via their SUSY loop corrections, and the 
process of up-down RGE evolution is iterated until a stable solution is 
obtained, using the complete set of two-loop RGEs.
A variety of non-universal soft SUSY breaking mass inputs are allowed,
including for this study independent gaugino masses.

\FIGURE[t]{\epsfig{file=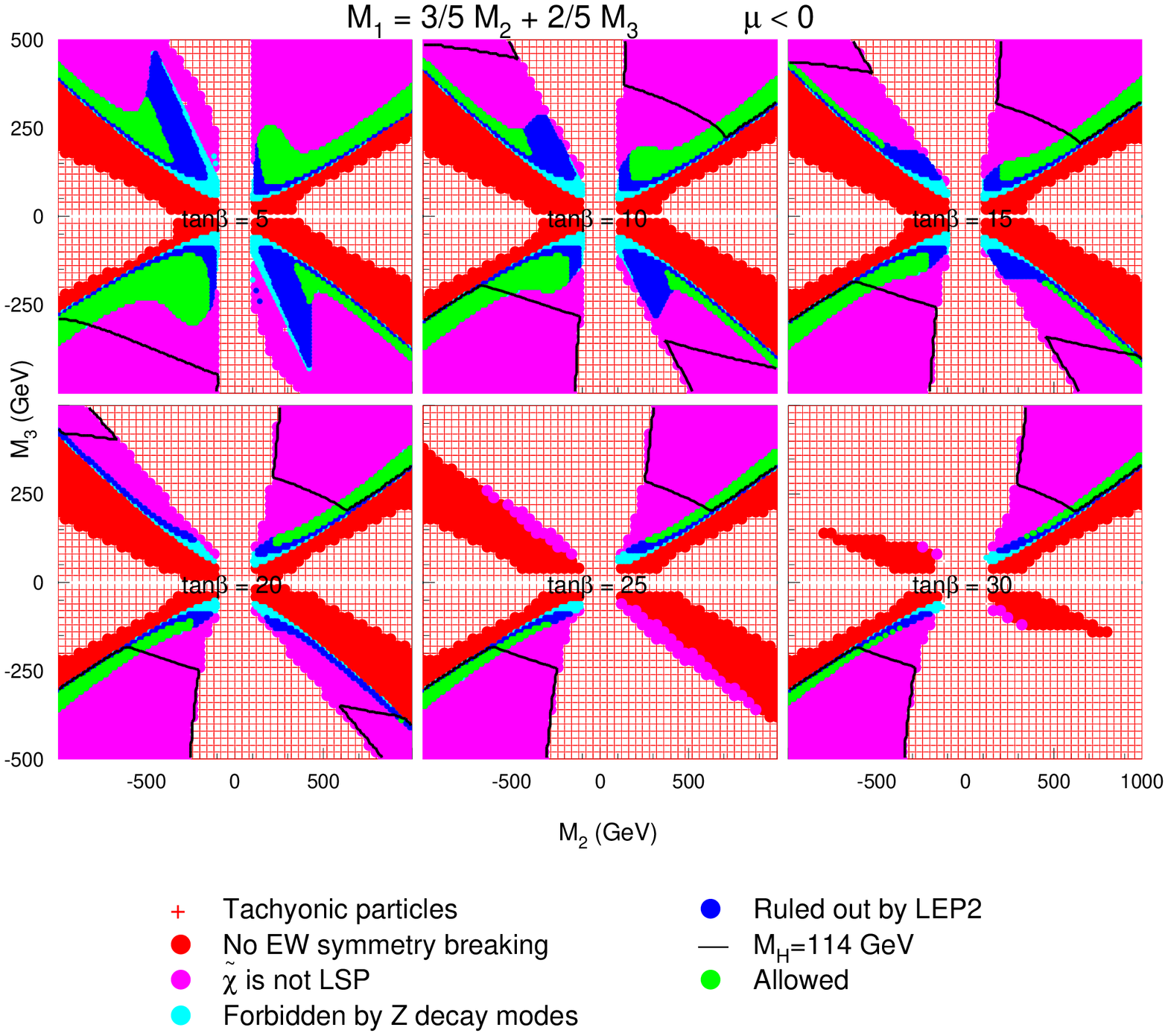,width=15cm} 
\caption{Allowed and excluded regions of the $M_2(M_c)$ vs. $M_3(M_c)$
         parameter plane in the Pati-Salam scenario for various $\tan\beta$
         values with $\mu <0$. Most of the parameter space is excluded by the 
         lack of REWSB (red shaded), tachyonic particles in the spectrum 
         (red crosses) or the stau being the LSP (magenta). Regions with
         viable spectra but in violation of limits from
         LEP (LEP2) are denoted by light blue (dark blue) shading.
         The black contour denotes regions where the light Higgs mass
         $m_h> 114$ GeV. Finally, the green shaded region is allowed
         by all constraints (except possibly those from LEP2 Higgs searches).}
\label{fig:PS1}}

Our first results are presented in Fig. \ref{fig:PS1}, where we 
show regions of the $M_2\ vs.\ M_3$ parameter plane for $\mu <0$ and
$\tan\beta = 5,\ 10,\ 15,\ 20,\ 25$ and $30$.
The regions with low values of $|M_i|$ denoted by red crosses give rise
to tachyonic sparticle masses. In these cases, the Yukawa coupling contribution
to the RGEs for top and bottom squarks is dominant, and quickly drives 
these squared masses to negative values. The red shaded regions do not give
rise to an appropriate radiative breakdown of EW symmetry. The magenta
regions give rise to a charged LSP, the lightest stau. This is the 
well-known problem associated with models of minimal gaugino mediation.
The light and dark blue shaded regions give rise to spectra in violation 
of LEP and LEP2 limits from sparticle searches. In the case of LEP bounds,
the $Z$ boson has non-negligible decays to sparticles, while in the LEP2 case
we require $m_{\tw_1}>100$ GeV, $m_{\tell_1}>92$ GeV, 
$m_{\tmu_1}>85$ GeV, $m_{\ttau_1}>68$ GeV and $m_{\tz_1}>37$ GeV\cite{lep2}.
The green shaded region obeys all constraints with the possible exception
of LEP2 Higgs searches. These range from $m_h>114$ GeV for low $\tan\beta$
to $m_h>91$ GeV for $\tan\beta >10$\cite{lep2higgs}. We show instead
the black contour where $m_h=114$ GeV, and where $m_h >114$ GeV occurs 
for larger values of $M_2$ and $M_3$. 

\FIGURE[t]{\epsfig{file=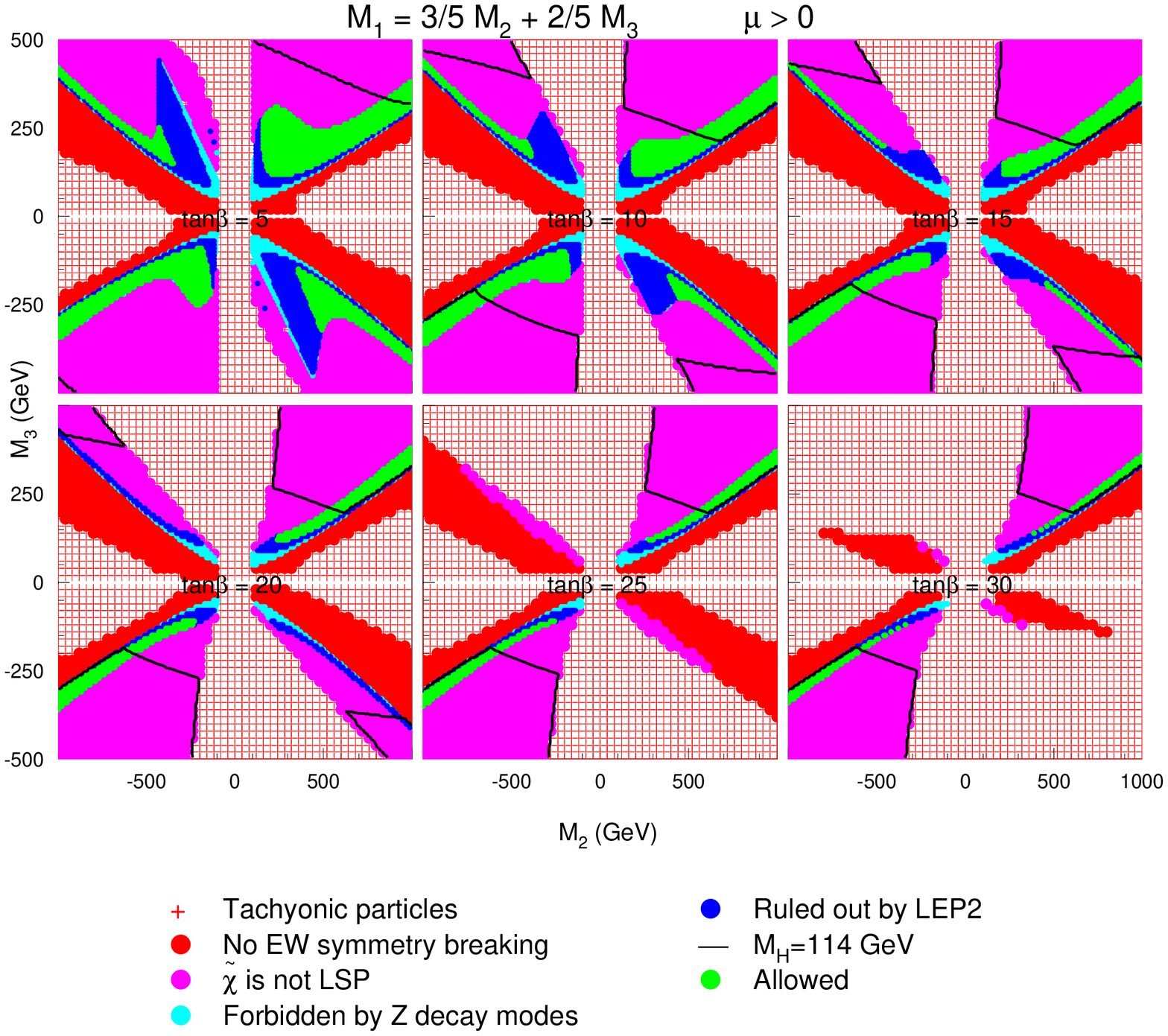,width=15cm} 
\caption{Same as in Fig. \ref{fig:PS1}, but with $\mu>0$.}
\label{fig:PS2}}

For $\tan\beta =5$, a significant green region survives all constraints
with the exception being that the light Higgs scalar has mass
$m_h<114$ GeV except for a tiny region in the lower left. 
As $\tan\beta$ increases, the green allowed region gradually shrinks, 
while the tachyonic and stau LSP regions increase.
This is reasonable in that as $\tan\beta$ increases, 
the $b$ and $\tau$
Yukawa couplings increase, and they contribute more strongly to
driving scalar squared masses to small or tachyonic values.
As $\tan\beta$ increases, however, the light Higgs mass $m_h$
also increases, giving rise to regions with viable
supersymmetric spectra.
This verifies the 
conjecture in Ref. \cite{so10in5d} that the increased flexibility in the 
non-universal gaugino masses can cure the stau LSP problem 
inherent in models of minimal gaugino mediation without beyond the GUT
scale running. However, we must note that the region of viable
parameter space is rather limited, and in fact disappears for
both high and low values of $\tan\beta$.

In our plots, both positive as well as negative gaugino mass
parameters are allowed. In fact, we observe a near symmetry of
parameter space when $M_2\to -M_2$ and $M_3\to -M_3$. In this case,
$M_1\to -M_1$ as well, and all gaugino masses flip sign, but 
remain with the same absolute value. The MSSM Lagrangian is in fact
invariant under the simultaneous sign 
change of the gaugino masses, the $A$ and $B$ parameters and 
$\mu$,\footnote{We thank X. Tata for emphasizing this point.}
so the near symmetry under sign flip of all gaugino masses may not
be surprising. If instead we flip the sign of just one of
$M_2$ or $M_3$, then of course there is much less symmetry, since
a different absolute value of the bino mass $M_1$ is generated.
Toward larger values of $\tan\beta$, 
this asymmetry increases because for $(M_2\cdot M_3)<0$, $|M_1|$ becomes
smaller than in the $(M_2\cdot M_3)>0$ case, and the tau Yukawa coupling 
can more easily drive the
right handed stau squared mass to negative values.

%A similar situation is shown in Fig. \ref{fig:PS2}, where we plot the same 
%parameters, but with $\mu >0$. 
%We observe the striking symmetry between the 
%opposite quadrants of the positive and negative 
%$\mu$ plots for each $\tan\beta$ 
%values. For example, the second quadrant of the $\tan\beta = 5$ frame in 
%Fig.\ref{fig:PS1} is essentially the same as the 180-degree rotated fourth 
%quadrant of the $\tan\beta = 5$ frame in Fig.\ref{fig:PS2}. This well 
%illustrates the invariance of the SUSY Lagrangian under 
% Although, in our case $B$ does not change 
%sign (and $A_0 = 0$ at $M_{GUT}$), 
%so this symmetry is only approximate, yet it 
%holds quite well, double checking our numerics.
%
%For a given sign of $mu$ there is a weaker symmetry between the opposite 
%quadrants for each $\tan\beta$. This caused by the sign change of the gaugino 
%masses, but not that of $B$ and $mu$. Finally, neighboring quadrants are 
%asymmetric, mainly because for a given $M_2$ ($M_3$) 
%value $|M_1|$ is different 
%for negative and positive $M_3$ ($M_2$). 

\begin{table}
\begin{center}
\caption{Representative weak scale (physical) sparticle masses 
(in GeV units) and parameters for three Pati-Salam scenarios. 
We take $M_c=1\times 10^{16}$ GeV.}
\bigskip
\begin{tabular}{lccc}
\hline
parameter & \multicolumn{3}{c}{value}  \\
\hline
$M_1(M_c)$        & 250   & 590    & 666 \\
$M_2(M_c)$        & 250   & 800    & 900 \\
$M_3(M_c)$        & 250   & 275    & 315 \\
$\tan\beta$       &   5   &  10    & 30  \\
$\mu$             & $>0$  & $<0$   & $>0$ \\
$\mu(M_{weak})$   & 337.6 &$-$153.5& 147.0 \\
$B(M_{weak})$     & 81.7  &$-$186.3& 62.5 \\
$m_{\tg}$         & 583.9 & 652.3  & 738.9 \\
$m_{\tu_L}$       & 519.5 & 729.1  & 823.6 \\
$m_{\td_R}$       & 502.8 & 541.6  & 614.0 \\
$m_{\tst_1}$      & 376.7 & 344.3  & 398.0 \\
$m_{\tb_1}$       & 485.2 & 536.3  & 565.5 \\
$m_{\te_L}$       & 174.0 & 531.4  & 596.6 \\
$m_{\te_R}$       & 101.4 & 222.1  & 249.6 \\
$m_{\tnu_{e}}$    & 156.1 & 525.5  & 591.3 \\
$m_{\ttau_1}$     & 98.4  & 215.4  & 162.6 \\
$m_{\tnu_{\tau}}$ & 156.0 & 524.2  & 577.7 \\
$m_{\tz_2}$       & 172.6 & 161.2  & 152.6 \\
$m_{\tw_1}$       & 171.3 & 153.2  & 144.7 \\
$m_{\tz_1}$       &  98.2 & 141.9  & 134.7 \\
$m_h      $       & 109.3 & 115.2  & 118.0 \\
$m_A      $       & 377.3 & 541.0  & 511.4 \\
$m_{H^+}  $       & 385.6 & 547.2  & 519.1 \\
\hline
\label{tab:PS}
\end{tabular}
\end{center}
\end{table}

To illustrate the sort of spectra that can occur in the PS scenario,
we show three cases in Table \ref{tab:PS} for low, 
moderate and high $\tan\beta$.
In the first case, {\it universal} gaugino masses are actually
taken along the line where $M_2=M_3$. Along this line, however, the
light Higgs mass $m_h$ is always below the limit from LEP2, and so
is excluded. In addition, for the first case, the $\ttau_1$ and $\tz_1$
are nearly mass degenerate.
The spectra for the other two cases
are taken from the upper right branches of allowable parameter
space in Figs. \ref{fig:PS1} and \ref{fig:PS2}.

Moving to regions beyond the black $m_h=114$ GeV contour, we find
the allowed
parameter space is bounded from below by regions where $\mu^2 <0$ (no REWSB)
and from above by where the $\ttau_1$ becomes the LSP. These cases 
typically
have small values of $|\mu|$, and hence a higgsino-like LSP. 
The $\tw_1$ ($\tz_2$) will decay into three body modes dominated by virtual
$W$ ($Z$) exchange. 
There is a very small mass gap between $\tw_1$ and $\tz_1$, 
so that $\tw_1$ and $\tz_2$ decay products will be quite soft, 
making collider searches challenging. 
One can expect that the search for lightest sleptons  -- 
selectron, smuon or stau --
will be a better choice to hunt for SUSY particles.

\FIGURE[t]{\epsfig{file=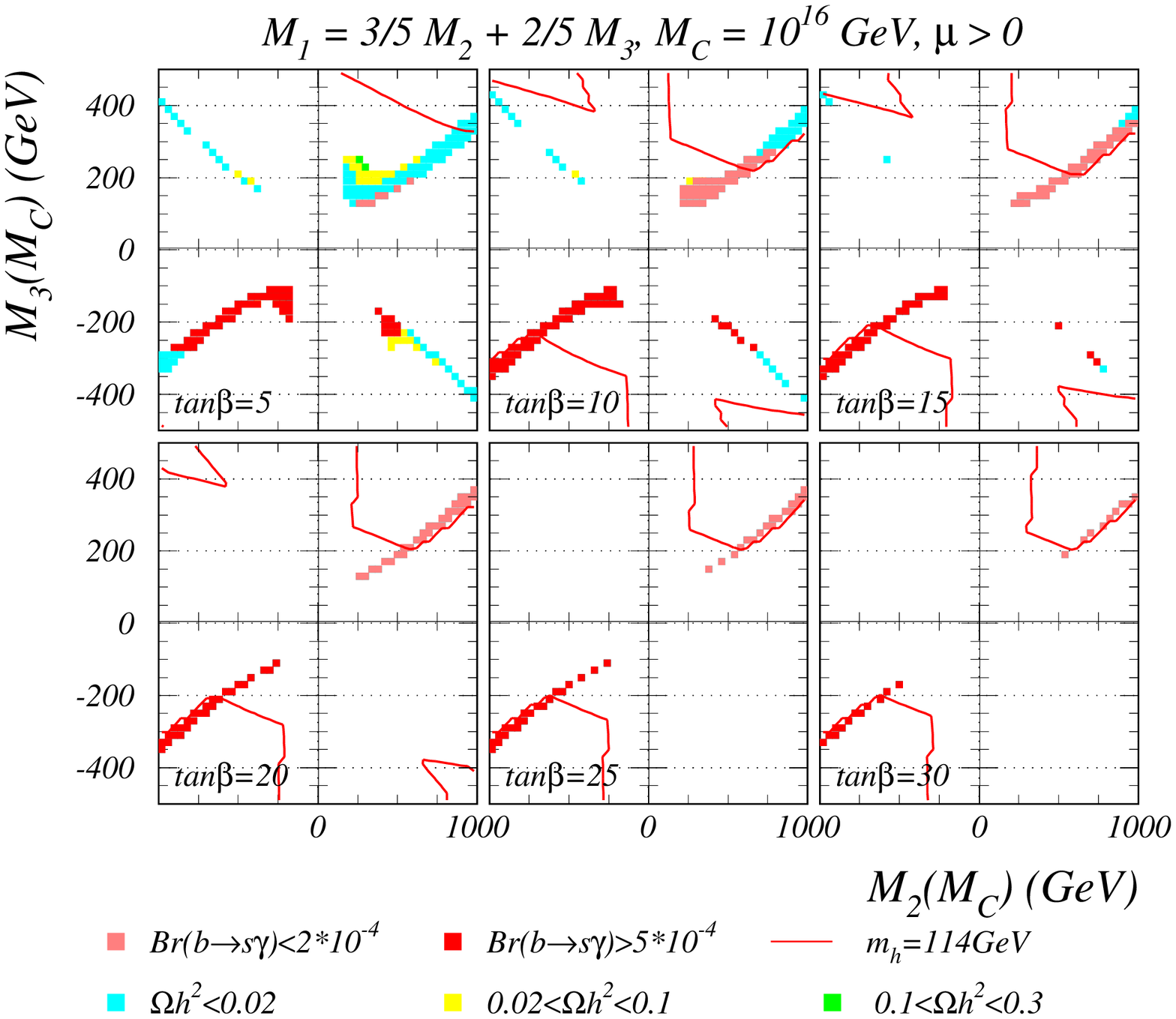,width=15cm} 
\caption{Same parameter space as in Fig. \ref{fig:PS2}, but 
showing the neutralino relic density $\Omega h^2$ in the 
viable regions of parameter space.}
\label{fig:PSrelic}}

\FIGURE[t]{\epsfig{file=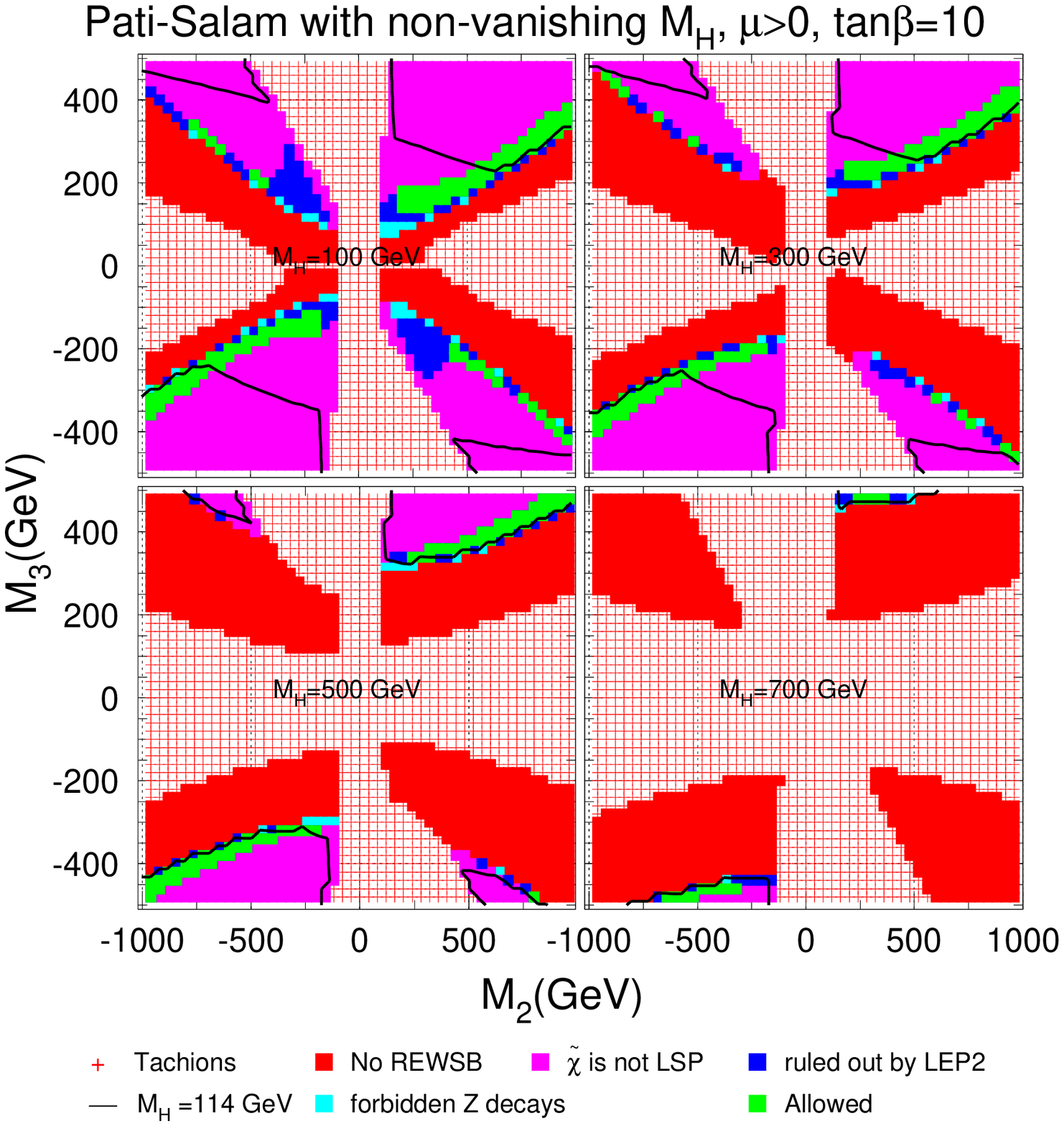,width=15cm} 
\caption{Allowed and excluded regions of the $M_2(M_c)$ vs. $M_3(M_c)$
         parameter plane for the Pati-Salam scenario with $\mu>0$
         and $\tan\beta =10$, but with
         $m_{H_u}(M_c)=m_{H_d}(M_c)=100$, 300 500 and 700 GeV.
         For $\mu < 0$ the result is qualitatively the same.}
\label{fig:PShiggs10}}

\FIGURE[t]{\epsfig{file=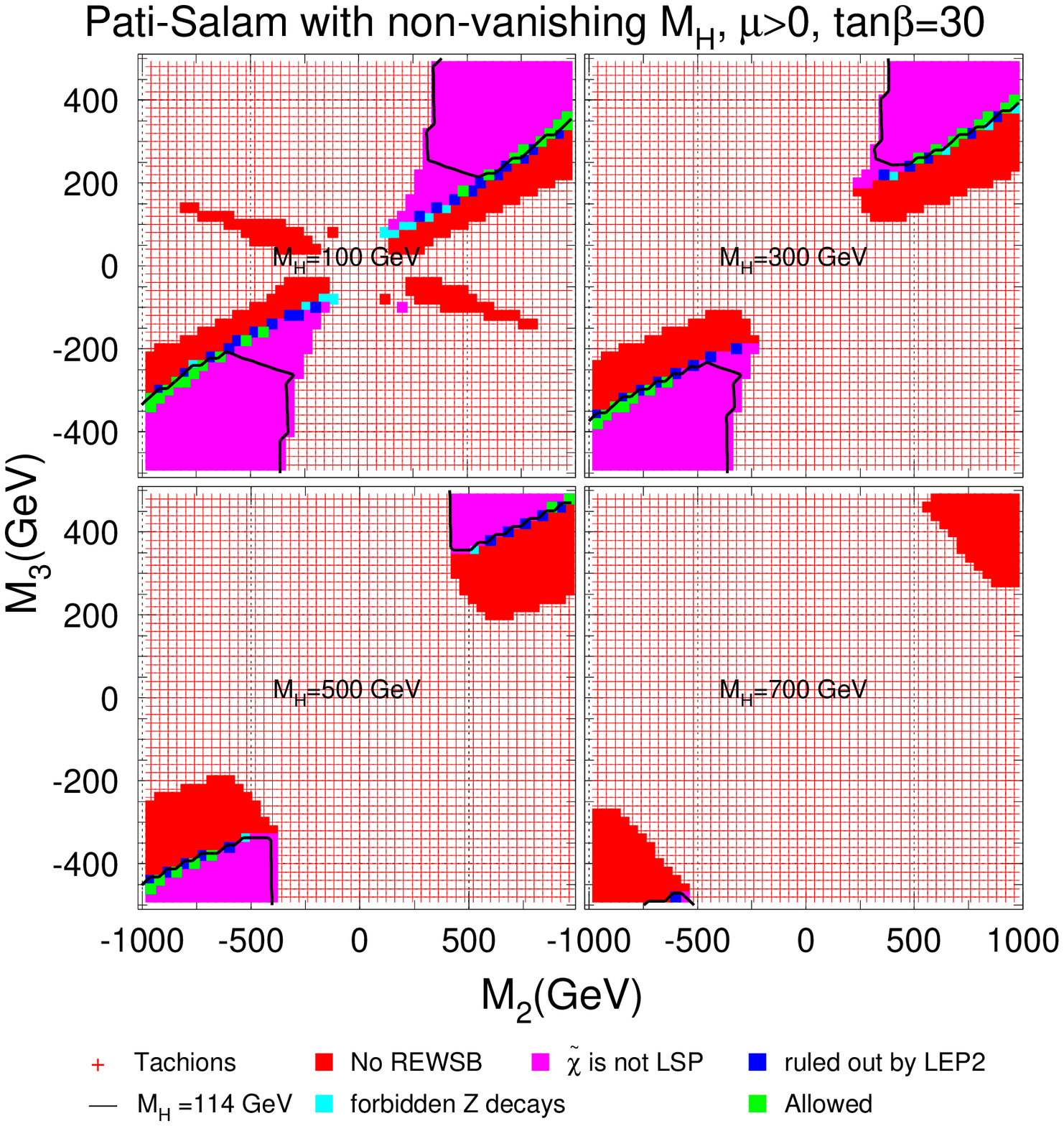,width=15cm} 
\caption{Allowed and excluded regions of the $M_2(M_c)$ vs. $M_3(M_c)$
         parameter plane for the Pati-Salam scenario with $\mu>0$
         and $\tan\beta =30$, but with
         $m_{H_u}(M_c)=m_{H_d}(M_c)=100$, 300 500 and 700 GeV.
         For $\mu < 0$ the result is qualitatively the same.}
\label{fig:PShiggs30}}

In Fig. \ref{fig:PSrelic}, we show in addition the relic density of
neutralinos and the region excluded by $BF(b\to s\gamma)$ constraints 
in the allowed parameter space regions from Fig. \ref{fig:PS2}.
For $BF(b\to s\gamma )$, we use the program of Ref. \cite{bsg},
upgraded to include running $b$-quark mass effects which are important
at large $\tan\beta$. 
The red regions give too high a value of $BF(b\to s\gamma)>5\times 10^{-4}$,
while pink regions give too low a value: $<2\times 10^{-4}$.
We also evaluate the muon anomalous magnetic moment $a_\mu$ following the
program of Ref. \cite{bbft}. All of the allowed region not excluded by 
$BF(b\to s\gamma )$ falls within the range
$-10\times 10^{-10}<a_\mu<60\times 10^{-10}$. 
%We caution the reader that, in view of the theoretical uncertainty 
%and the imminent new experimental analysis for $a_\mu$, 
%and also potential contributions from small off-diagonal SSB 
%masses to $BF(b\to s\gamma)$, these excluded regions should be interpreted 
%as illustrative.
We caution the reader that the  excluded regions we present here and
later on due to the $BF(b\to s\gamma )$ and $a_\mu$ criteria  should be
interpreted as illustrative. The first reason for this is the
theoretical uncertainty and the imminent new experimental analysis
for $a_\mu$. Secondly, the results on $BF(b\to s\gamma)$ are quite
model-dependent --- the potential contributions from small
off-diagonal SSB masses to $BF(b\to s\gamma)$ can easily work in
both ways --- they can easily increase or decrease the deviations
of $BF(b\to s\gamma)$ from the SM value.

For the neutralino relic density, we use the recent 
calculation of Ref. \cite{bbb}, 
which includes all relevant co-annihilation channels as well as
relativistic thermal averaging of annihilation cross sections 
times neutralino velocity. The blue regions have
$\Omega h^2<0.02$, while yellow regions have $0.02<\Omega h^2 <0.1$.
The current cosmologically favored amount of CDM
is $0.1<\Omega h^2 <0.3$ (green). 
Thus, the neutralino relic density in the PS
scenario is typically quite low compared with the cosmologically 
favored values of the CDM density. This is easy to understand, since
the allowed parameter space of the PS scenarios have an LSP with 
a large higgsino component. In this case, there is a large
annihilation rate into $WW$, $ZZ$ and $Zh$ pairs, which results
in a general reduction of the expected CDM density. Since our scenario is a
higher dimensional model, the low neutralino relic density can easily be 
augmented by other forms of CDM related to extra dimensions.
Thus, the predicted low neutralino relic density does not restrict the allowed 
parameter space in the PS case, and leaves room for other forms of CDM
inherent in extra-dimensional models. 

If the Higgs fields of an extra dimensional SUSY GUT propagate in the bulk,
then they may gain non-negligible soft SUSY breaking masses as well.
This additional freedom of non-zero Higgs squared masses at the 
compactification scale may allow additional parameter space to open up
in the PS scenario.
In Fig. \ref{fig:PShiggs10}, we plot the allowed parameter
space in the $M_2\ vs.\ M_3$ plane for $\tan\beta =10$, but this 
time allowing for non-zero soft SUSY breaking Higgs masses
$m_{H_u}=m_{H_d}=100$, 300, 500 and 700 GeV. In fact as the
Higgs masses increase, the allowed parameter space moves to
higher absolute $M_3$ values, but also diminishes in area. The general
diminution of parameter space occurs because in the RGEs for
squared slepton masses, there exist terms where Higgs masses multiply Yukawa
couplings. These contributions drive the soft SUSY breaking slepton masses
to lower values than in the $m_{H_u}=m_{H_d}=0$ case, giving larger
regions of parameter space with a stau LSP.

Since the diminution of soft breaking sfermion masses is due to
RGE terms involving Yukawa couplings, we expect the effect to be amplified for
large $\tan\beta$, where the $\tau$ Yukawa coupling gets large.
In fact this is shown in Fig. \ref{fig:PShiggs30}, where we show the
same results, but for $\tan\beta =30$. 
In this case, there is very little allowed parameter space for low
values of Higgs masses, and no parameter space left on the frames shown 
for the very large values of $m_{H_i}$ ($i=u,\ d$).

\section{Gaugino mediation with independent gaugino masses}
\label{sec:general}

\FIGURE[t]{\epsfig{file=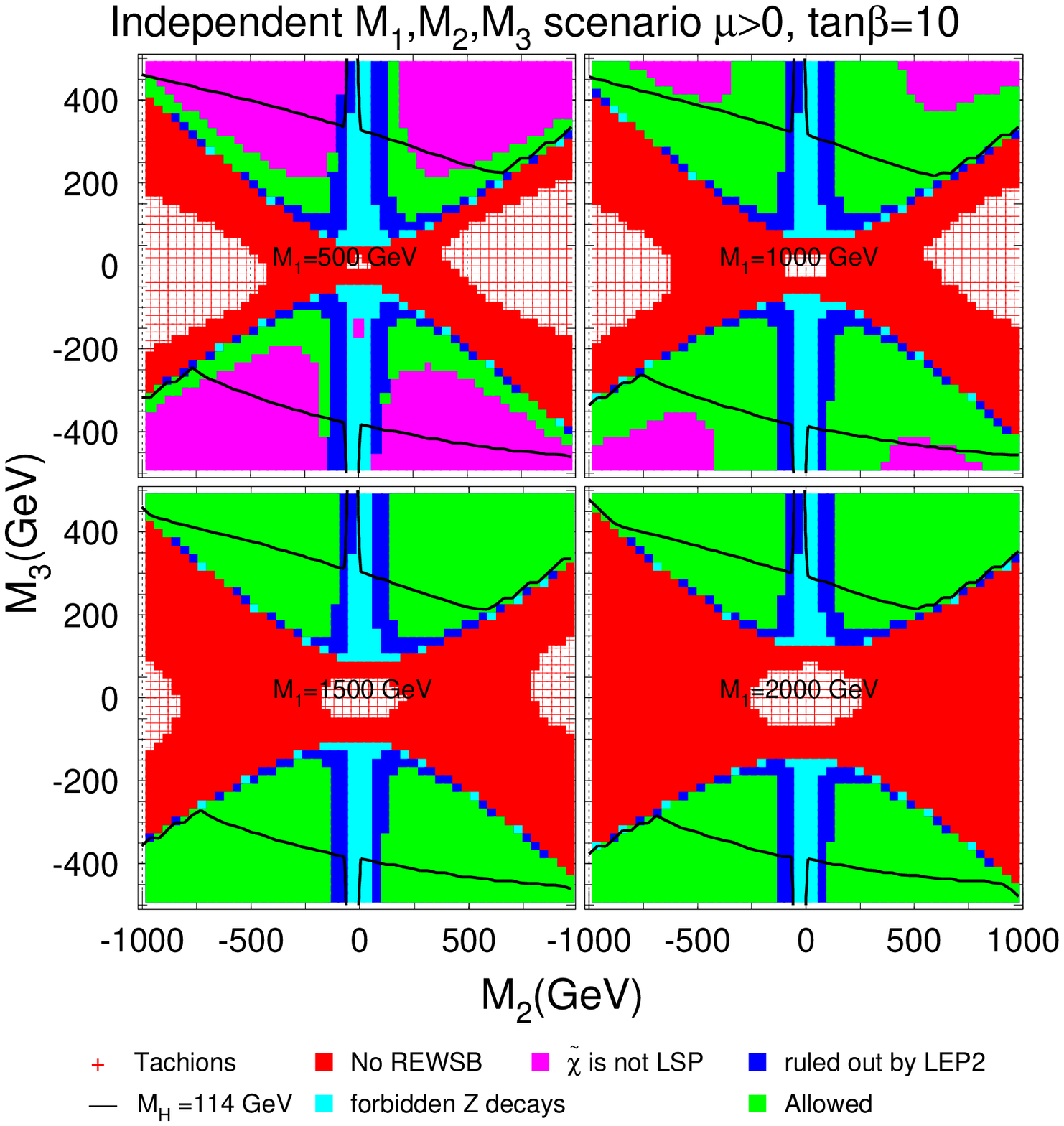,width=15cm} 
\caption{Allowed and excluded regions of the $M_2(M_c)$ vs. $M_3(M_c)$
         parameter plane for independent gaugino masses at $M_c$, 
         $\tan\beta=10$ and $\mu>0$. The four frames are shown for 
         $M_1(M_c)=500$ 1000, 1500 and 2000 GeV. 
         For $\mu < 0$ the result is qualitatively the same.}
\label{fig:gen10}}

A more general scenario occurs when all three gaugino masses
are independent at the compactification scale, as in Eq. (\ref{eq:genM}).
In this case, we again scan the $M_2\ vs.\ M_3$ plane, but this time for
different independent fixed values of gaugino mass $M_1$.
In Fig. \ref{fig:gen10}, we show the parameter space for $\tan\beta =10$,
$\mu >0$ and fixed values of $M_1=500$, 1000, 1500 and 2000 GeV. The shading 
of the different forbidden and allowed regions is the same as in
Figs. \ref{fig:PS1} and \ref{fig:PS2}. We see that for $M_1=500$
GeV, for low values of $M_3$, the parameter space is forbidden
either by tachyonic particles or no REWSB; for small values of
$M_2$, additional parameter space is ruled out by
constraints from LEP2 on the mass of the lightest chargino.
As $M_1$ increases, the $U(1)_Y$ contribution to the running of 
the soft SUSY breaking squared mass $m^2_{\ttau_R}$ increases, so that the
region of parameter space excluded by a $\ttau_1$ LSP shrinks. The
allowed region of parameter space grows to become a major fraction
of the $M_2\ vs.\ M_3$ plane.

\FIGURE[t]{\epsfig{file=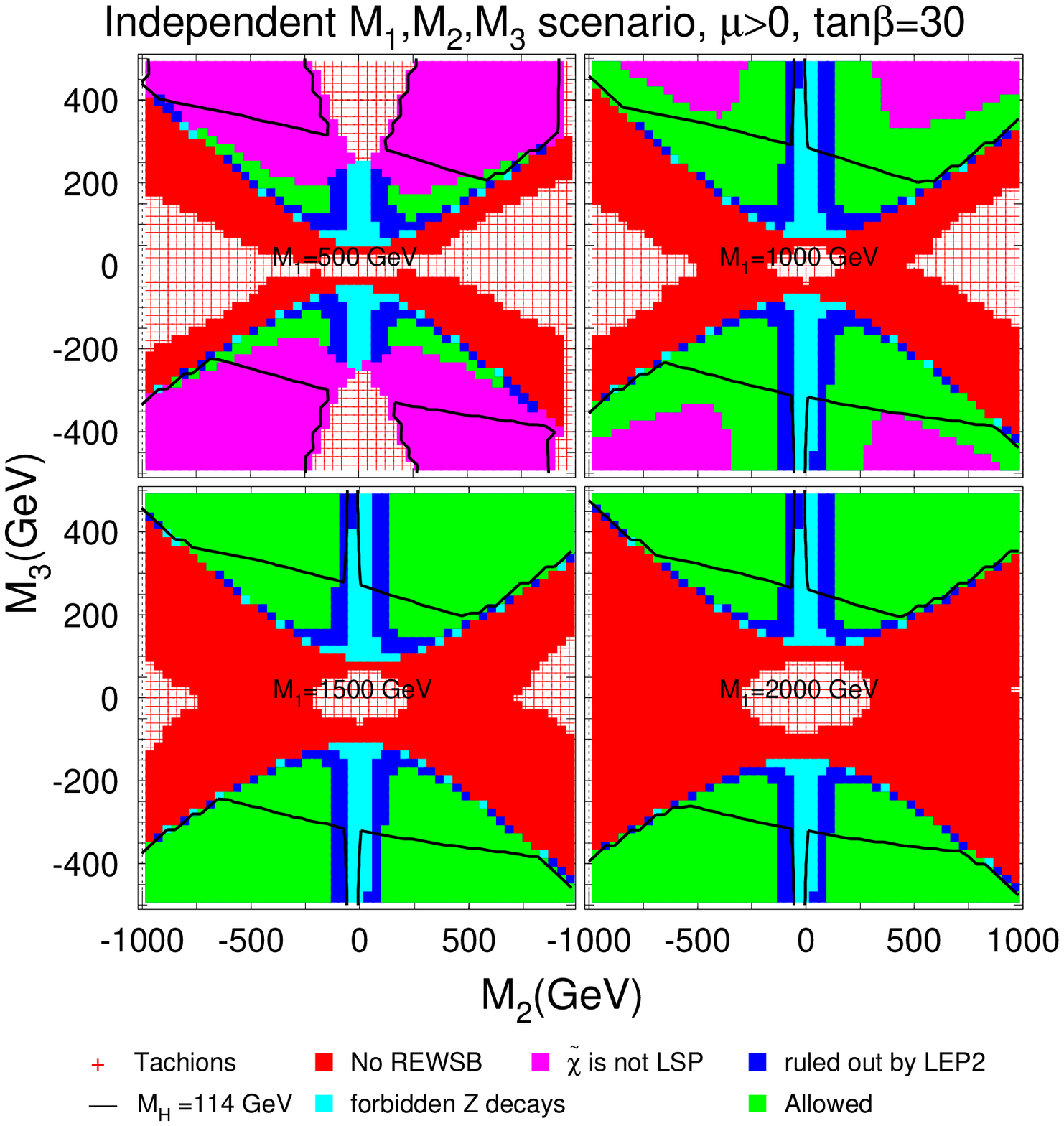,width=15cm} 
\caption{Allowed and excluded regions of the $M_2(M_c)$ vs. $M_3(M_c)$
         parameter plane for independent gaugino masses at $M_c$, 
         $\tan\beta=30$ and $\mu>0$. The four frames are shown for 
         $M_1(M_c)=500$, 1000, 1500 and 2000 GeV. }
\label{fig:gen30}}

\FIGURE[t]{\epsfig{file=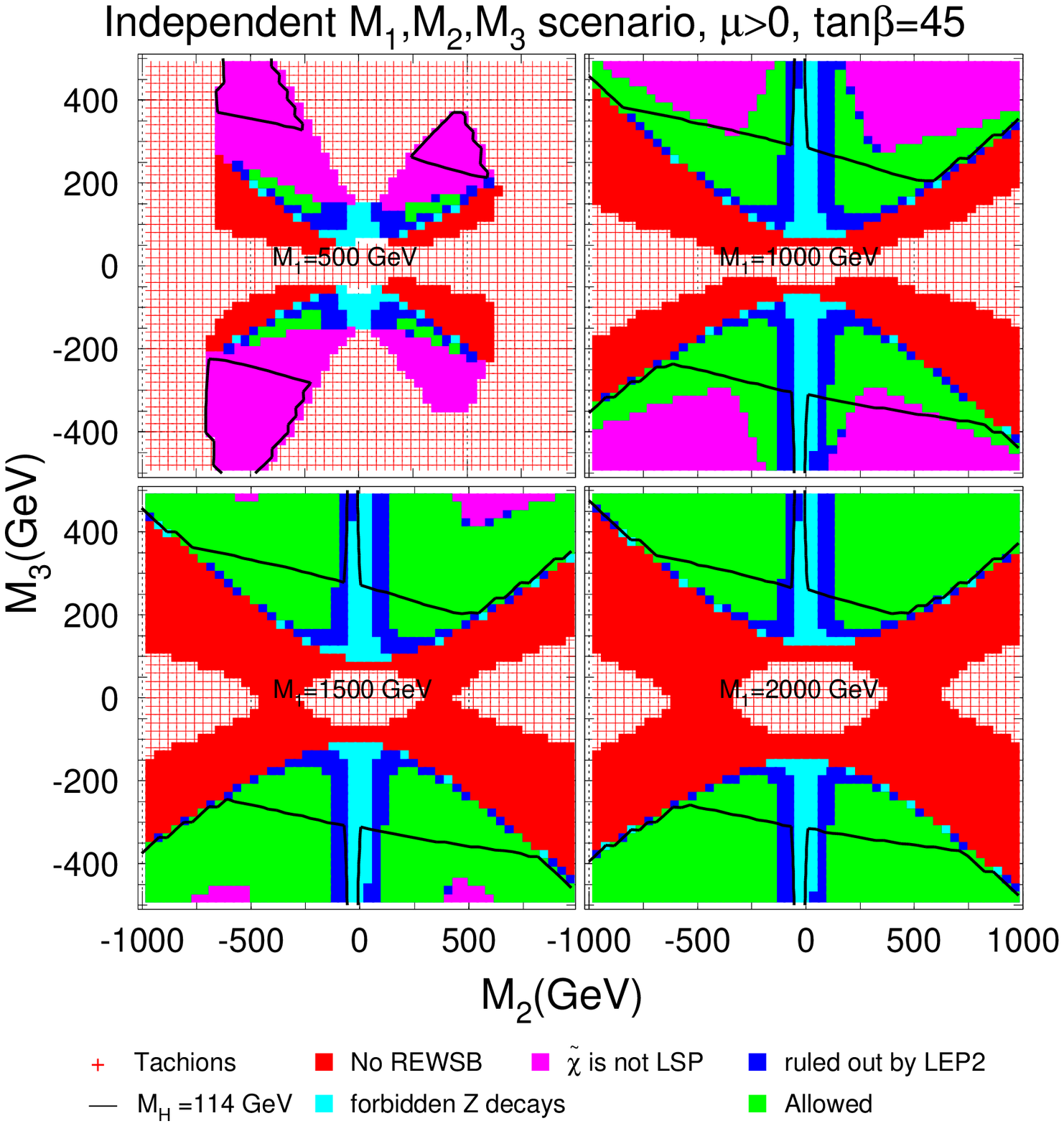,width=15cm} 
\caption{Allowed and excluded regions of the $M_2(M_c)$ vs. $M_3(M_c)$
         parameter plane for independent gaugino masses at $M_c$, 
         $\tan\beta=45$ and $\mu>0$. The four frames are shown for 
         $M_1(M_c)=500$, 1000, 1500 and 2000 GeV. }
\label{fig:gen45}}

We also show the same $M_2\ vs.\ M_3$ parameter space for fixed 
values of $M_1$ for larger $\tan\beta$ values of 30 and 45 in 
Figs. \ref{fig:gen30} and \ref{fig:gen45}, respectively.
As $\tan\beta$ increases, the $\tau$ Yukawa coupling also increases,
which suppresses the stau masses, so that more of parameter
space is disallowed by either the $\ttau_1$ LSP or tachyonic
constraints. However, in all cases, if $M_1$ is taken to be sufficiently 
large, then the gauge contribution to $\ttau_i$ running wins out, and
much of the parameter space with large gaugino masses becomes again allowed. 

\begin{table}
\begin{center}
\caption{Representative weak scale (physical) sparticle masses 
(in GeV units) and parameters for three independent gaugino mass scenarios. 
We take $M_c=1\times 10^{16}$ GeV.}
\bigskip
\begin{tabular}{lccc}
\hline
parameter & \multicolumn{3}{c}{value}  \\
\hline
$M_1(M_c)$ & 500 & 500 & 1500 \\
$M_2(M_c)$ & 150 & 1000 & 500 \\
$M_3(M_c)$ & 300 & 350  & 400 \\
$\tan\beta$ & 10 & 10 & 30  \\
$\mu$ & $>0$ & $>0$ & $>0$ \\
$M_2(M_{weak})$ & 112.0 & 796.1 & 393.2 \\
$\mu (M_{weak})$ & 392.4 & 181.6 & 466.4 \\
$B(M_{weak})$ & 42.1 & 244.4 & 22.1 \\
$m_{\tg}$   & 688.9 & 815.2 & 903.6 \\
$m_{\tu_L}$ & 595.0 & 909.6 & 824.2 \\
$m_{\td_R}$ & 596.3 & 672.9 & 787.4 \\
$m_{\tst_1}$& 471.6 & 424.5 & 645.7 \\
$m_{\tb_1}$ & 556.4 & 666.6 & 729.4 \\
$m_{\te_L}$ & 139.1 & 653.2 & 425.2 \\
$m_{\te_R}$ & 189.8 & 189.1 & 555.0 \\
$m_{\tnu_{e}}$ & 114.4 & 648.4 & 417.7 \\
$m_{\ttau_1}$ & 128.4 & 176.7 & 398.7 \\
$m_{\tnu_{\tau}}$ & 113.6 & 646.9 & 398.2 \\
$m_{\tz_2}$ & 206.6 & 186.3 & 470.7 \\
$m_{\tw_1}$ & 104.1 & 178.2 & 364.5 \\
$m_{\tz_1}$ & 103.7 & 156.2 & 362.9 \\
$m_h      $ & 114.1 & 117.8 & 118.3 \\
$m_A      $ & 404.8 & 665.9 & 546.9 \\
$m_{H^+}  $ & 412.7 & 671.1 & 554.1 \\
\hline
\label{tab:gen}
\end{tabular}
\end{center}
\end{table}

In Table \ref{tab:gen}, we show several representative cases from the
allowed parameter space with independent gaugino masses.
The first case has $M_1(M_c)=500$ GeV, while $M_2(M_c)=150$ GeV
and $M_3(M_c)=300$ GeV. The most notable feature of this model is
that $M_2(M_{weak})=111.9$ GeV, compared with a $\mu$ value of
$392.4$ GeV. This results in the LSP being wino-like, and the
$\tw_1$ and $\tz_1$ being nearly mass degenerate. In fact, in this case 
the $\tw_1$ is just beyond the kinematic reach of LEP2. However, its
dominant decay mode is $\tw_1\to \pi^+\tz_1$, and since the
$\tw_1 -\tz_1$ mass gap is very small, the pion will be soft and not
easily detected. The chargino is sufficiently long-lived
($c\tau\simeq 0.1$ cm) that there may be a detectable
track followed by kink in chargino production and decay.
This is similar to the case from minimal anomaly-mediated SUSY breaking 
models (mAMSB)\cite{anomalyMSB1,anomalyMSB2}, which also has a wino-like LSP. 
Other intriguing features of the spectra are that the light Higgs $h$
is just near the LEP2 bound, and that the sneutrinos are the lightest
of all the sleptons, and that the left sleptons are far lighter than
right sleptons, in contrast to the case in the minimal supergravity (mSUGRA)
model.

The second case shown in Table \ref{tab:gen} is quite different
in that now $M_1$ and $\mu$ are comparable at the weak scale, and 
the LSP is a higgsino-bino mixture. There is
a substantial mass gap between $\tw_1$ and $\tz_1$, and the right-
sleptons are lighter than left-sleptons. The $\ttau_1$ is quite light,
and in fact the decays of $\tw_1$ and $\tz_2$ are dominantly into 
staus.

The third case in Table \ref{tab:gen} is at large $M_1$ and 
large $\tan\beta$, and the LSP is again wino-like, as in case 1.
The lighter sleptons are dominantly left eigenstates.

\FIGURE[t]{\epsfig{file=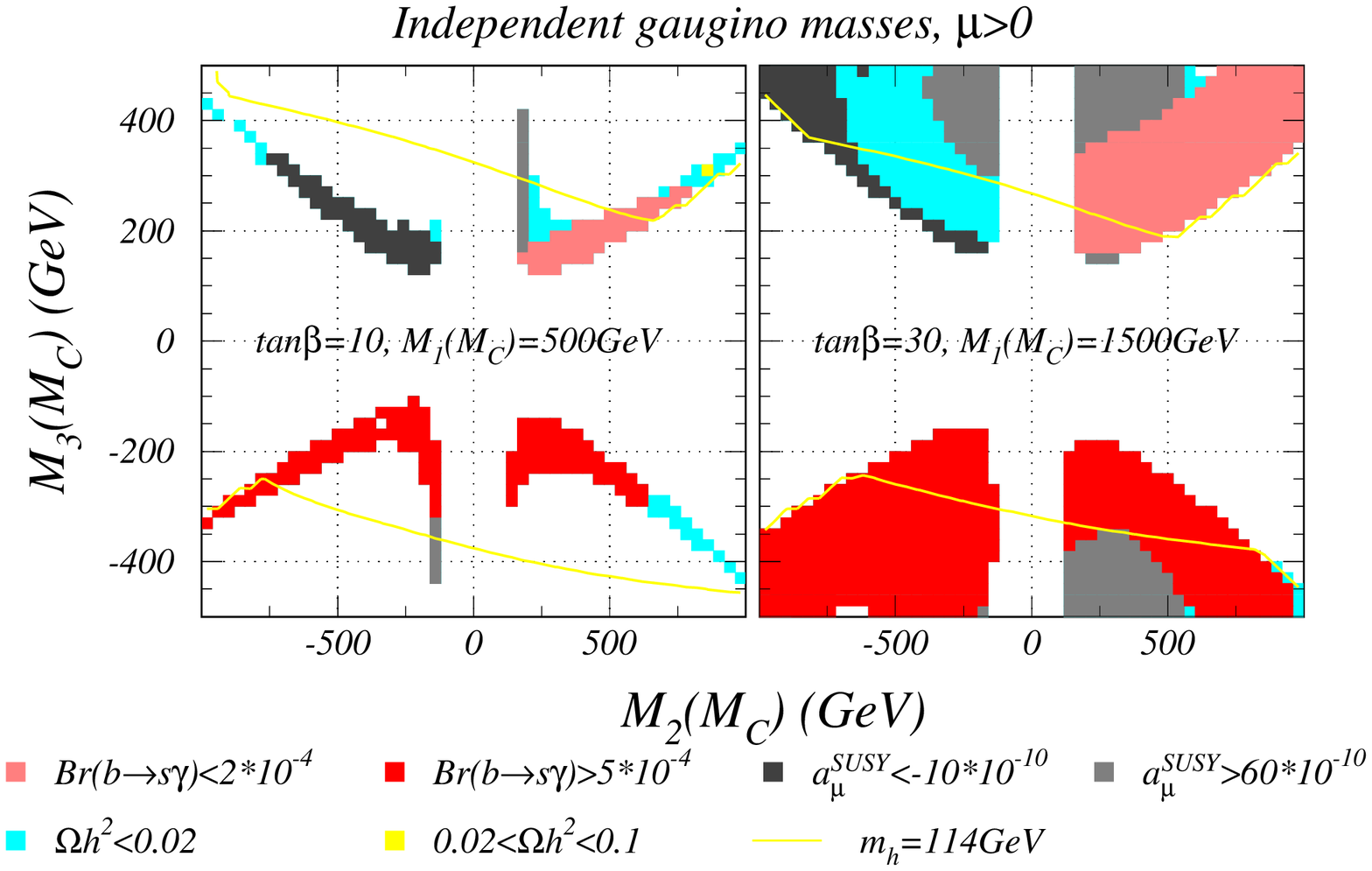,width=15cm} 
\caption{Regions of neutralino relic density in models with
independent gaugino masses for $\tan\beta =10$ and $M_1=500$ GeV, 
and for $\tan\beta =30$ and $M_1=1500$ GeV. We take
$\mu >0$.}
\label{fig:genrelic}}

In Fig. \ref{fig:genrelic}, we show the relic density of neutralinos,
the $BF(b\to s\gamma )$ excluded region, and the region excluded by
$a_\mu$ constraints,
for two sample frames of the case with general gaugino masses. The first
frame has $M_1=500$ GeV, $\tan\beta =10$ and $\mu >0$. 
The color coding is similar to Fig. \ref{fig:PSrelic}, 
except that now regions excluded by $a_\mu$ are also shown as light 
and dark gray regions. When a region is excluded by both 
$BF(b\to s\gamma)$ and $a_\mu$ it is colored red or pink.
We note that the excluded regions are {\it not} 
symmetric in $BF(b\to s\gamma )$ under $M_3\to -M_3$, since
the $A$-parameters evolve quite differently, giving rise to different
stop masses and also $\mu$ parameter.

Throughout almost
all the allowed parameter space, the relic density $\Omega h^2$ is less
than 0.02 (blue), {\it i.e.} not enough to explain the CDM needed by
cosmological observations. 
This is not surprising in that over much of parameter space, 
the LSP is either wino-like or higgsino-like. In the first case, 
there is a large co-annihilation rate between $\tz_1$ and $\tw_1$ 
that leads to a small relic density.
In the second case, there is a large annihilation rate into $WW$, $ZZ$
and $Zh$ pairs, which also suppresses the relic density.
There are a few points for large 
positive $M_2$ and $M_3$ where $\Omega h^2>0.02$. In this region, 
$\mu$ and $M_2$ are comparable at the weak scale, and the mass
gap between $\tw_1$ and $\tz_1$ can reach up to 20 GeV, thus suppressing the
$\tw_1-\tz_1$ coannihilation rate, and increasing the relic density.
The second frame of Fig. \ref{fig:genrelic} shows the relic density
for $M_1=1500$ GeV, $\tan\beta =30$ and $\mu >0$. In this case, the relic
density is low over all the parameter space shown, due to the 
presence of a wino-like LSP. Again, a low neutralino relic density leaves
room for other potential forms of CDM and does not restrict the allowed 
parameter space of the model.

\section{Conclusions}
\label{sec:concl}

Gaugino-mediated SUSY breaking is especially attractive in that it
leads to a natural solution to the problem of flavor-changing and $CP$
violating processes that are generic to the MSSM. 
If the soft SUSY breaking boundary conditions from gaugino mediation
apply at the GUT scale, then models with universal gaugino masses lead
to a charged (stau) LSP. Additional running of soft SUSY breaking parameters
above the GUT scale can lift the stau mass to the level where a 
neutralino LSP is possible. However, if the running occurs in a simple GUT, 
then the values of $\tan\beta$ consistent with Yukawa unification
may conflict with the predicted value of $\tan\beta$ from minimal
gaugino mediation with a small $B$ term. 

In this paper we examined an 
alternative solution to the stau LSP problem, in that we allow non-universal
gaugino masses. Extra dimensional GUT models can be easily constructed
which include gaugino-mediated SUSY breaking, but non-universal
gaugino masses. The additional freedom from the non-universal gaugino masses
allows us to find viable regions of model parameter space where 
SUSY spectra are generated consistent with constraints on non-tachyonic 
sparticles, REWSB and constraints from LEP2. We examined the case of
two independent gaugino masses, which occurs when SUSY breaking takes place on
a hidden sector brane with Pati-Salam symmetry. In this case, limited regions
of viable parameter space were found, usually with a higgsino-like LSP.
Those regions vanish for the high values of $\tan\beta\sim 45-55$
where one could expect Yukawa coupling unification. 

We also examined the case where all three gaugino masses were independent.
In these models, choosing $M_1$ large enough opens up large
regions of viable parameter space. In this case, the LSP might be either
wino-like or higgsino-like. Regions with large $\tan\beta$ are allowed
provided one chooses $M_1$ to be sufficiently large. 
We illustrated regions of parameter space constrained by muon $g-2$
and $BF(b\to s\gamma )$, although these constraints must be interpretted
with caution.
Due to the nature of the LSP, the relic
density of neutralinos in these models is generally below expectations 
from cosmology; in this case, some other type of matter 
(axions or hidden sector
states) would be needed to account for the bulk of CDM needed in the universe. 

\section*{Acknowledgments}
 
This research was supported in part by the U.S. Department of Energy
under contract number DE-FG02-97ER41022.
A.M. is supported in part by the National Science Foundation
under Grant PHY-0071054, and by a Research Innovation Award from
Research Corporation.
R.D. thanks S. Raby and Y. Nomura for discussions. R.D. is supported in
part by DOE grant DOE/ER/01545-830.
	
% ---- Bibliography ----
%

\end{document}